\documentclass[aps,twocolumn]{revtex4-1}

\usepackage[utf8]{inputenc}
\usepackage[T1]{fontenc}
\usepackage{textcomp}
\usepackage[english]{babel}
\usepackage{graphicx}
\usepackage{latexsym,amsmath,amssymb,amsthm}
\usepackage{makeidx}
\usepackage[usenames,dvipsnames]{color}
\usepackage[unicode=true,colorlinks=true,linkcolor=RoyalBlue,citecolor=RoyalBlue]{hyperref}
\usepackage[table]{xcolor}
\usepackage{natbib}
\usepackage{enumerate}
\usepackage{ulem}

\newcommand{ \Vq }{ {\mathbf{q}} }
\newcommand{ \rp }{ {\Vq} }
\newcommand{ \drp }{ {\dot{\rp}} }
\newcommand{ \ddrp }{ {\ddot{\rp}} }
\newcommand{ \Rp }{ {Q} }

\newcommand{ \PsitRp }{ {\Psi_{ \Theta , \Rp }} }
\newcommand{ \EVMC }{ {E_{\text{VMC}}} }
\newcommand{ \Lagr }{ {\mathcal{L}} }
\newcommand{ \massp }{ {m} }
\newcommand{ \dPsi }{ {\dot{\Psi}} }

\newcommand{ \VF }{ {\mathbf{F}} }
\newcommand{ \grad }{ {\frac{\partial}{\partial \rp_i}} }
\newcommand{ \gradH }{ {\frac{\partial H}{\partial \rp_i}} }
\newcommand{ \Oo }{ {\mathcal{O}} }

\newcommand{ \erfc }[1]{ \text{erfc}\left( #1 \right) }
\newcommand{ \Vol }{ \mathcal{V} }
\newcommand{ \Htwo }{$\text{H}_2$ }

\newcommand{ \Catm }{ C_{\text{atm}} }
\newcommand{ \phipw }{ \phi^{\text{pw}} }
\newcommand{ \phiatm }{ \phi^{\text{atm}} }
\newcommand{ \phibat }{ \phi^{\text{bat}} }
\newcommand{ \uyuk }{ \text{u}_{\text{\tiny YUK}} }

\makeindex

\begin{document}
	
\title{Two-Dimensional Hydrogen Structure at Ultra-High Pressure}

\author{Francesco \surname{Calcavecchia}}
\email{francesco.calcavecchia@gmail.com}
\affiliation{LPMMC, UMR 5493 of CNRS, Universit{\'e} Grenoble Alpes, 38042 Grenoble, France}

\author{Thomas D. K\"uhne}
\affiliation{Dynamics of Condensed Matter, Department of Chemistry, University of Paderborn, Warburger Str. 100, D-33098 Paderborn, Germany}
\affiliation{Paderborn Center for Parallel Computing and Institute for Lightweight Design, Department of Chemistry, University of Paderborn, Warburger Str. 100, D-33098 Paderborn, Germany}

\author{Markus \surname{Holzmann}}
\email{markus.holzmann@grenoble.cnrs.fr}
\affiliation{LPMMC, UMR 5493 of CNRS, Universit{\'e} Grenoble Alpes, 38042 Grenoble, France}
\affiliation{Institut Laue Langevin, BP 156, F-38042 Grenoble Cedex 9, France}

\begin{abstract}
We introduce a novel method that combines the accuracy of Quantum Monte Carlo simulations with ab-initio Molecular Dynamics, in the spirit of Car-Parrinello. 
This method is then used for investigating the structure of a two-dimensional layer of hydrogen at $T=0~\text{K}$ and high densities.
We find that metallization is to be expected at $r_s \approx 1.1$, with an estimated pressure of $1.0\cdot10^3~a_0~\text{GPa}$, changing from a graphene molecular lattice to an atomic phase.
\end{abstract}

\maketitle

Hydrogen is considered the holy grail of high pressure physics.
The origin of this saying is the prediction made by Wigner and Huntington in 1935 \cite{wigner:764}, suggesting the possibility that 
above the density corresponding to $r_s=1.63$, hydrogen could turn into a metallic solid with a bcc atomic structure. 
Here, the density $\rho$ is parametrized by $r_s=a/a_B$ where $a$ is the mean inter-particle distance and $a_B$ is the Bohr radius.
Wigner and Huntington estimated the necessary pressure to attain metallization to be of the order of $25~\text{GPa}$.
In 1967, based on BCS theory, Ashcroft argued that such a phase would be superconductive at room temperature \cite{1968PhRvL..21.1748A}.

These predictions aroused the interest of the scientific community and challenged high-pressure physics to produce metallic hydrogen in the laboratory. 
After the introduction of diamond anvil cells \cite{RevModPhys.55.65}, it became apparent that $25~\text{GPa}$ is insufficient for the metallization of hydrogen. 
Nowadays experimentalists are able to compress hydrogen by applying pressures of more than $300~\text{GPa}$ \cite{eremets:dense_hydrogen,PhysRevLett.108.125501}, revealing a surprisingly rich variety of phases \cite{silvera:pathway_hydrogen,Nellis:wh_hydrogen_review,RevModPhys.84.1607,Diaseaal1579}.


Experimental advances in high pressure hydrogen also triggered the development of new computational methods based on quantum Monte Carlo calculations \cite{PhysRevB.36.2092,PhysRevLett.93.146402,PhysRevLett.100.114501,CalcavecchiaKuehne:solid-hydrogen-ASWF}. These allow  quantitative comparisons and provide more accurate theoretical predictions for pressures out of  experimental reach.
Here, we present a first quantum Monte Carlo study of $2$-D hydrogen layers at zero temperature. 
Two-dimensional hydrogen is especially interesting for two reasons.
First, it may open a new pathway to metallic or even superconducting hydrogen.
Secondly, its comprehension might yield crucial clues for understanding the new $3$-D phase formed of alternating layers \cite{PhysRevB.85.214114}.

For the investigation of the structure of $2$-D hydrogen, we introduced a novel algorithm based on Variational Monte Carlo (VMC) \cite{McMillan:1965uq}.
VMC is able to treat quantum correlation effects while remaining computationally affordable, and it can take full advantage of High Performance Computers, since it is embarrassingly parallel.
Moreover, as we will see, the use of a variational principle on a chosen Ansatz, allows us to investigate the localization of electrons, providing additional information about hydrogen properties.


Our algorithm can be schematized in two layers:
\begin{enumerate}
   \item 
   At the core of our simulations we employed the VMC method to calculate the electronic Born-Oppenheimer energy surface.
   For the sake of simplicity, we only considered classical protons, enabling us to treat the Coulomb potential generated by $N$ protons as a static external potential. 
   Therefore, given the protonic positions $\Rp\equiv({\bf q}_1,\dots {\bf q}_N)$ and a trial wave function $\PsitRp$ with $n$ variational parameters $\Theta\equiv(\theta_1,\dots \theta_n)$, the VMC energy reads as
   \begin{equation}
      \EVMC(\Theta,\Rp) = \frac{\langle \PsitRp \mid H \mid \PsitRp \rangle}{\langle \PsitRp \mid \PsitRp \rangle},
   \end{equation}
   where the Hamiltonian $H$ accounts for the electronic kinetic energy and the Coulomb potential energy.
   \item On top of VMC, we used an optimization algorithm for finding the $\Theta$ and $\Rp$ which minimize $\EVMC$ to find the zero temperature ground state structures at different densities.
\end{enumerate}

In the following, the latter optimization method is illustrated. For that purpose, 
we begin by introducing a Lagrangian for our system
\begin{equation}
   \Lagr = \frac{1}{2} \sum_{i=1}^{N} \massp \drp_i^2 + \mu \langle \dPsi \mid \dPsi \rangle - \langle \Psi \mid H - \Lambda \mid \Psi \rangle \label{eq:lagrangian}
\end{equation}
where $\massp$ is the protonic mass, $\mu$ is a fictitious mass, and $\Lambda$ is a Lagrange multiplier which ensures the normalization condition \mbox{$\langle \Psi \mid \Psi \rangle = 1$} of the trial wave function $|\Psi \rangle$.

The corresponding Euler-Lagrange equation for the protons 
leads to 
\begin{eqnarray}
   \massp \ddrp_i &=&\VF_i 
   \nonumber
   \\
    \VF_i &=&
    - \grad \left( \langle \Psi \mid H \mid \Psi \rangle - \Lambda \langle \Psi \mid \Psi \rangle + \mu \langle \dPsi \mid \dPsi \rangle  \right) \, , \label{eq:motion_ions}
\end{eqnarray}
where $\VF_i$ is the force acting on the protons.
Such an equation is related to Born-Oppenheimer molecular dynamics (MD).
In fact, the latter is recovered by setting $\mu=0$ and assuming that $\Psi$ is an exact eigenstate, or very close to it, so that the forces can be evaluated using the Hellmann-Feynman theorem
\begin{equation}
   \VF_i^{BO} = - \langle \Psi \mid \gradH \mid \Psi \rangle
\end{equation}

The dynamics of the electronic wave function for fixed protons is entirely contained in the time-dependence of the variational parameters 
\begin{equation}
  \dot{\Psi}(\Theta)  =  \sum_{i=1}^n  \dot{\theta}_i \Oo_i \Psi(\Theta) \, ,
\end{equation}
where $\Oo_i \equiv \partial \log  \Psi /\partial \theta_i-\langle \partial \log  \Psi /\partial \theta_i  \rangle $, and $\langle \cdot \rangle \equiv \langle \Psi \mid \cdot \mid \Psi \rangle$. The resulting Euler-Lagrange equation for the $\Theta$ variables
then gives
\begin{equation}
   \mu \sum_{i=1}^n \ddot{\theta}_i \langle  \Oo_i \Oo_j  \rangle = - \langle \Oo_j (H - \Lambda) \rangle, \label{eq:motion_theta}
\end{equation}
or
\begin{equation}
   \mu\ddot{\theta}_i =  {\cal F}_i,
\end{equation}
where  ${\cal F}_i$ acts as a generalized force on the electronic parameters
\begin{equation}
{\cal F}_i= - \sum_j \langle \Oo_i \Oo_j \rangle^{-1} \langle \Oo_j (H - \Lambda) \rangle. \label{eq:SR}
\end{equation}
Note that Eq.~\ref{eq:motion_ions} and \ref{eq:motion_theta} represents a coupled electron-ion dynamics that can be utilized to facilitate quantum Monte Carlo based \textit{ab-initio} molecular dynamics in the spirit of Car-Parrinello molecular dynamic (CPMD) \cite{PhysRevLett.55.2471,PhysRevLett.98.066401}, which keeps the electronic degrees of freedom very close to the instantaneous ground state. However, the noise in the nuclear forces computed in this way needs to be compensated by means of a modified Langevin equation \cite{PhysRevB.73.041105,PhysRevLett.98.066401,PhysRevLett.100.114501}. 

In this work, we have focused on this set of equations for minimizing the variational energy and finding the optimal protonic structure.
We have neglected the dynamic, and simply used the forces to reach the minimum,
similar to the Stochastic Reconfiguration method\cite{PhysRevB.71.241103} as described in \cite{paper:calcavecchia_kuehne}.

The accuracy of VMC depends on the quality of the underlying trial wave function \cite{Pierleoni200889,PhysRevB.91.115106}. 
Here, we have considered basic Jastrow-Slater (JS) trial wave functions \cite{paper:calcavecchia_kuehne},
\begin{equation}
\begin{split}
   \Psi(R)  & = \det \phi_n({\bf r}_i) e^{- \sum_{i,j} \uyuk(r_{ij})},~\text{where} \\
   \uyuk(r) & \equiv A \; \frac{1 - e^{-Fr}}{r}. 
\end{split}
\end{equation}
This is to say that a Yukawa form for the electron-electron and electron-proton pair correlation
accounts for symmetrical correlations ($A$ and $F$ are variational parameters, $R=({\bf r}_1, \dots {\bf r}_N)$ the electronic coordinates) is employed here. For the Slater determinant, we have used four different kinds of orbitals, $\phi_n({\bf r})$:
\begin{description}
   \item[plane-waves] $ \phipw(\mathbf{r},\mathbf{k}) \equiv \exp(-i \mathbf{k} \cdot \mathbf{r}) $, where $\mathbf{k}$ labels the Fermi k-vectors;
   \item[atomic] $ \phiatm(\mathbf{r},\mathbf{q}) \equiv \exp(-\Catm || \mathbf{q} - \mathbf{r} ||) $, where $\Catm$ is a variational parameter;
   \item[bi-atomic] $ \phibat(\mathbf{r},\mathbf{q}_1,\mathbf{q}_2) \equiv \phiatm(\mathbf{r},\mathbf{q}_{1}) + \phiatm(\mathbf{r},\mathbf{q}_{2}) $, where $\mathbf{q}_1$ and $\mathbf{q}_2$ belong to the same \Htwo molecule;
   \item[DFT] orbitals resulting from a DFT calculation employing the PBE exchange-correlation energy functional and the bare Coulomb pseudo-potential.
\end{description}

\begin{figure}[htbp]
  \centering
    \includegraphics[width=0.45\textwidth]{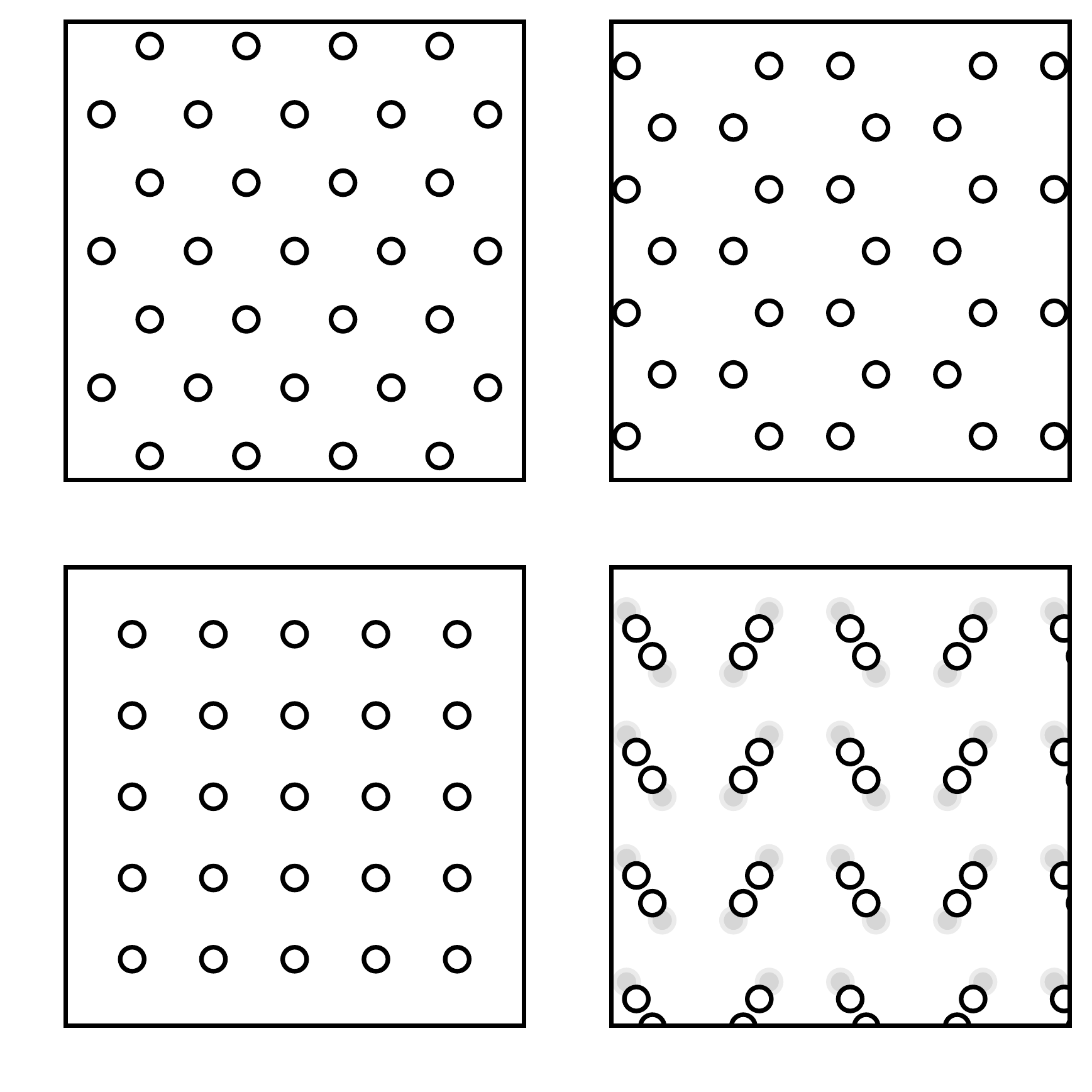}
  \caption{2D protonic configurations considered in this work. Starting from the bottom-left panel, in clock-wise order: Squared, triangular, atomic graphene-like, molecular graphene-like. In the molecular graphene-like structure, light grey markers are used to illustrate the connection to the atomic structure.}
  \label{fig:2DStructures_structures}
\end{figure}

We applied this novel algorithm as described above for investigating the high pressure structures of a 2D layer of hydrogen, where point-like protons are strictly confined to a plane but electrons can move in all three dimensions. 
We considered four configurations for the protons, forming either a square, triangular, or (atomic or molecular) graphene-like lattice, as illustrated in Fig.~\ref{fig:2DStructures_structures}. 
Our periodic systems contained $N=128$ hydrogen atoms for the triangular lattice and $N=144$ for all of the other structures. 
Twist averaged boundary conditions were applied to reduce finite size effects \cite{PhysRevE.64.016702,PhysRevB.94.035126}, which are know to be particularly important in the metallic phase. 

\begin{figure}[htbp]
  \centering
    \includegraphics[width=0.45\textwidth]{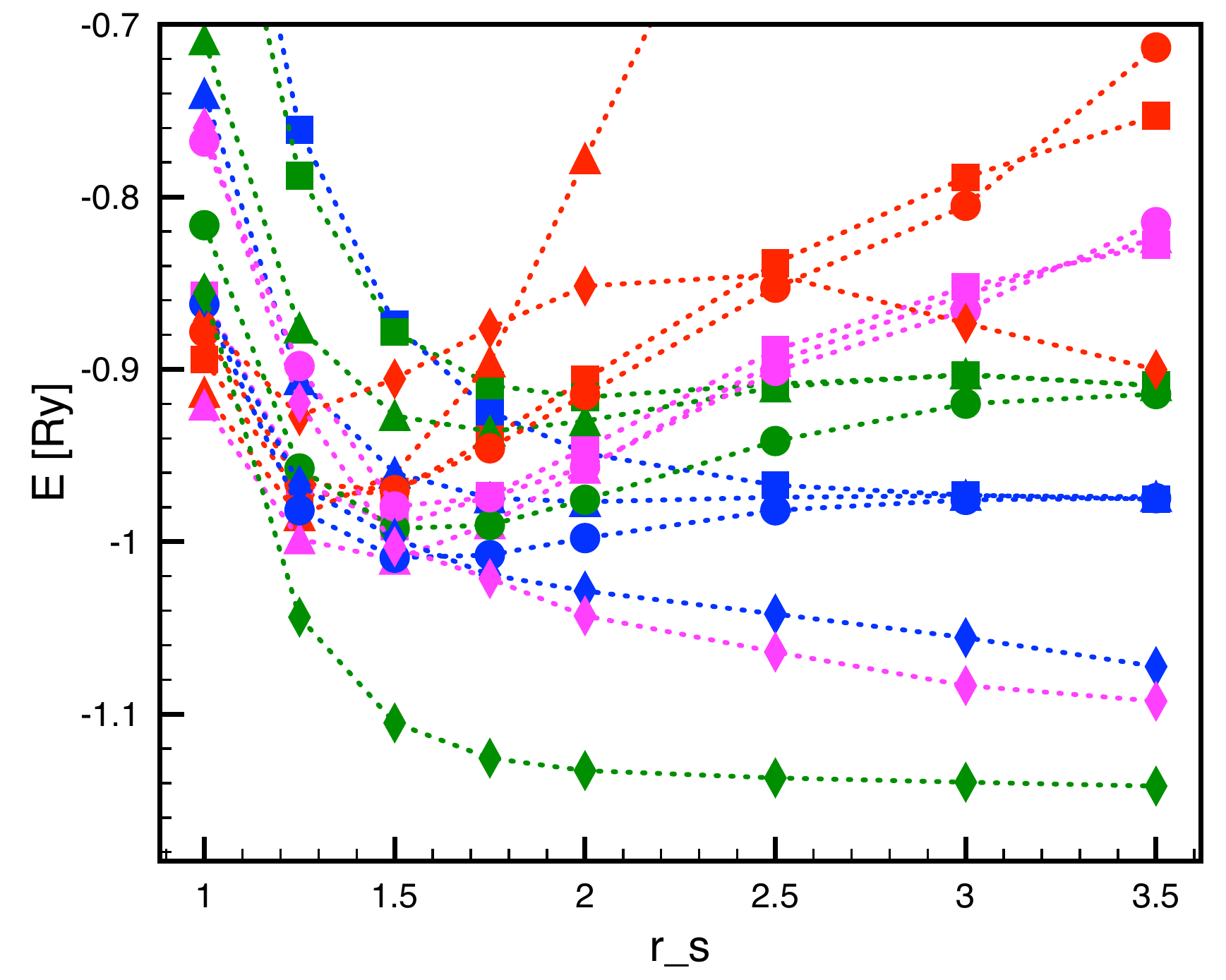}
  \caption{Variational energies obtained by optimizing the wave function variational parameters. We used a symbol code for labeling different proton configurations (square=squared, triangle=triangular, circle=atomic graphene-like, rhombus=molecular graphene-like) and a color code for the employed trial wave function (red=plane waves, blue=atomic, green=bi-atomic, magenta=DFT).}
  \label{fig:WfOpt_energies}
\end{figure}

For each of these structures, we first computed the variational energy, optimizing only the wave function parameters. 
The resulting energies corresponding to densities in between $1 \le r_s=a/a_B \le 3.5$, where $a=(\pi \rho)^{-1/2}$ is the mean inter-particle distance in 2D, are reported in Fig.~\ref{fig:WfOpt_energies} and Table \ref{energytable}. 
Out of the investigated structures, the molecular graphene-like lattice structure with bi-atomic orbitals turned out the most stable one at low density/pressure whereas the triangular atomic lattice with DFT orbitals becomes favorable at high densities, $r_s  \lesssim 1.1$. 
At this level, energies are rather widespread depending on both the considered lattice structure and on the underlying orbitals used in the Slater determinant.

\begin{figure}[htbp]
  \centering
    \includegraphics[width=0.45\textwidth]{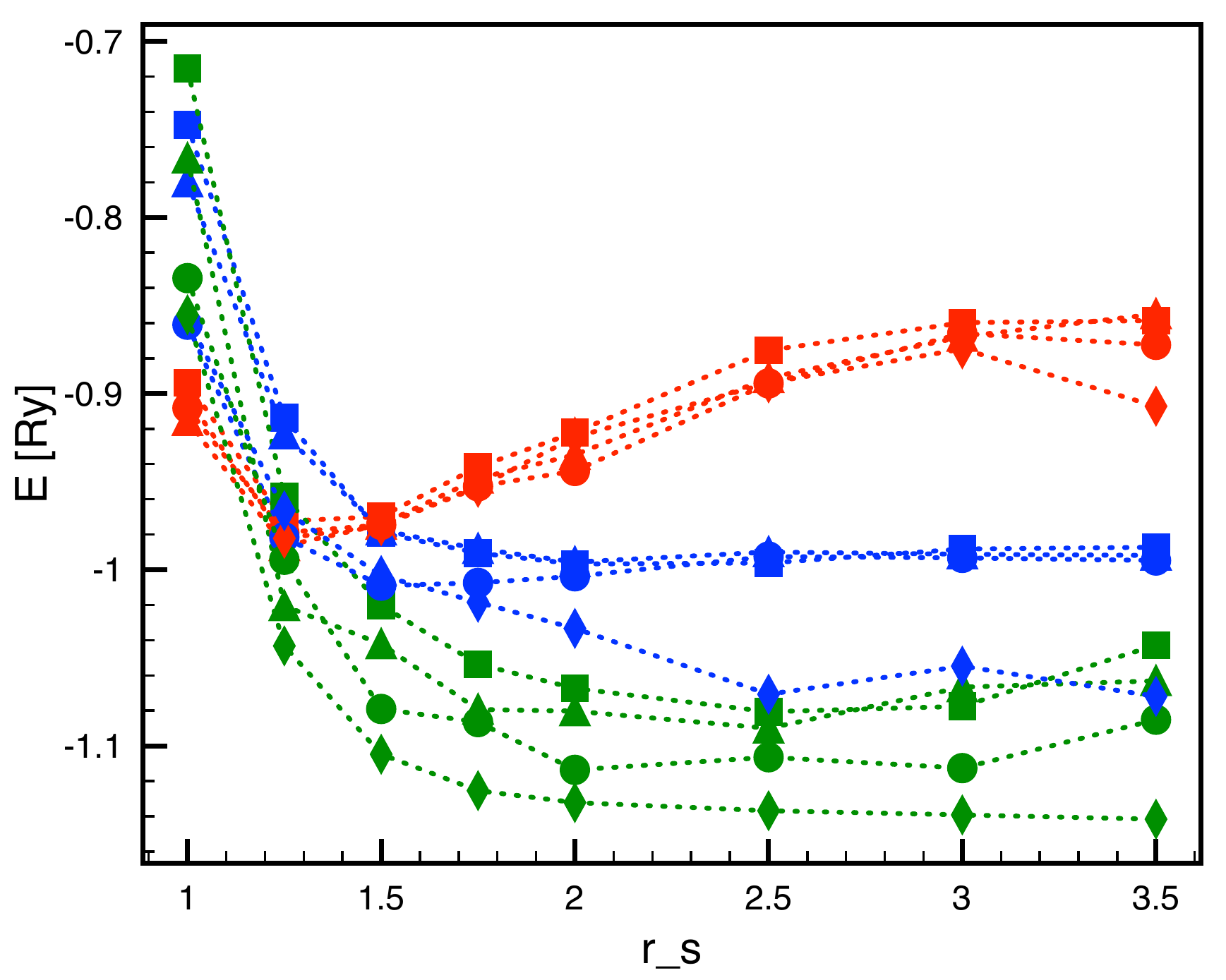}
  \caption{Variational energies obtained by optimizing the wave function variational parameters and the proton positions. We used a symbol code for labeling different proton configurations (square=squared, triangle=triangular, circle=atomic graphene-like, rhombus=molecular graphene-like) and a color code for the employed trial wave function (red=plane waves, blue=atomic, green=bi-atomic).}
  \label{fig:StructureRelaxation_energies}
\end{figure}

\begingroup
\begin{table*}
\centering
\begin{tabular}{|cc||c|c|c|c|c|c|c|c|}
\hline
 $r_s$ & & $1.0$ &$1.25$ & $1.5$  & $1.75$ &$2.0$ &$2.5$ & $3.0$ &$3.5$  \\
 \hline
\hline
square & DFT & -0.8574(13) &  -0.965(1) & -0.9910(7) & -0.9734(5) &  -0.9459(5) & -0.8888(5) & -0.8521(4) &-0.8278(6) \\
& pw  & -0.8943(3) & -0.9731(4) & -0.9709(3) & -0.9403(4) & -0.9058(2) & -0.8384(4) & -0.7886(5) & -0.7528(9) \\
& rel.  & -0.8938(5) & -0.9725(3) & -0.9698(2) & -0.9417(4) & -0.9220(2) & -0.8754(3) & -0.8597(4) & -0.8585(6) \\
& atomic &  -0.4894(4) & -0.7610(3) & -0.8740(2) &  -0.9244(1) & -0.94873(8) & -0.96719(3) & -0.97291(2) &  -0.97537(2) \\
& rel. & -0.7474(5) &  -0.9137(3) & -0.9791(3) & -0.9905(2) & -0.9967(2) & -0.9961(2) & -0.98824(7) & -0.98723(5) \\
& bi-atomic & -0.5429(5) & -0.7876(4) & -0.8783(3) &  -0.9092(1) & -0.9162(2) & -0.9085(3) & -0.9034(3) & -0.9093(4) \\
& rel. & -0.7153(6) & -0.9584(2) & -1.0203(3) & -1.0536(3) & -1.0674(3) & -1.0806(3) & -1.0776(3) &  -1.0428(4) \\
\hline
triangle & DFT & \cellcolor{green} -0.9209(12) & -0.9984(5) & -1.0103(4) & -0.9897(6) & -0.9576(4) & -0.8956(4) & -0.8565(5) & -0.8239(6) \\
& pw  & -0.9136(6) & -0.9850(5) & -0.9605(9) & -0.8959(9) & -0.778(1) & -0.5513(9) & -0.4939(5) & -0.3997(4) \\
& rel. &  \cellcolor{orange} -0.9154(3) & -0.9865(3) & -0.9744(4) & -0.9482(4) & -0.9347(3) & -0.8911(4) & -0.8677(2) & -0.8548(4) \\
& atomic &  -0.7403(4) & -0.9062(3) & -0.9601(2) & -0.9747(2) & -0.97684(8) & -0.97423(5) & -0.97327(2) & -0.97390(1) \\
& rel. &   -0.7799(5) & -0.9230(5) & -0.9777(3) & -0.9881(3) & -0.9957(2) & -0.98994(8) & -0.99120(7) & -0.99188(5) \\
& bi-atomic &  -0.7091(3) & -0.8762(2) & -0.9270(2) & -0.9360(1) & -0.9300(1) & -0.9103(2) & -0.9028(3) & -0.9097(3) \\
& rel. &   -0.7664(4) & -1.0205(3) & -1.0422(4) & -1.0793(2) & -1.0803(2) & -1.0901(3) & -1.0667(5) & -1.0631(2) \\
\hline
graphene-a & DFT & -0.7679(44) & -0.898(3) & -0.980(2) & -0.974(1) & -0.9566(8) & -0.9009(9) & -0.8658(5) & -0.8146(9) \\
& pw & -0.8784(4) & -0.9662(4) & -0.9700(4) & -0.9458(3) & -0.9148(3) & -0.8527(7) & -0.8051(7) & -0.713(1) \\
& rel. & -0.9082(3) & -0.9789(3) & -0.9745(3) & -0.9526(4) & -0.94390(3) & -0.8942(3) & -0.8659(5) & -0.8723(7) \\
& atomic &  -0.8622(3) & -0.9817(2) & -1.0092(2) &  -1.0077(2) & -0.9979(1) & -0.98176(7) & -0.97570(5) & -0.97473(2) \\
& rel. & -0.8609(4) & -0.9816(3) & -1.0090(3) & -1.0075(2) & -1.0037(1) & -0.9925(1) & -0.9933 & -0.99476(9) \\ 
& bi-atomic &  -0.8164(5) & -0.9575(2) & -0.9923(2) & -0.9903(1) & -0.9757(1) & -0.9416(1) & -0.9199(2) & -0.9143(3) \\
& rel. &   -0.8344(5) &  -0.9943(3) & -1.0791(2) & -1.0865(2) & -1.1137(1) & -1.1065(3) & -1.1127(2) & -1.0852(3) \\
\hline
graphene-m & DFT & -0.7598(30) & -0.919(3) & -1.003(1) & -1.021(1) & -1.043(1) & -1.0640(8) & -1.0834(6) & -1.0924(8) \\
& pw & -0.8720(4) & -0.9270(4) & -0.9055(6) & -0.8761(5) & -0.8517(6) & -0.8517(7) & -0.873(1) & -0.900(1) \\
& rel. & -0.9024(4) & -0.9822(3) & -0.9750(3) &  -0.9532(3) & -0.9247(4) & -0.8939(4) & -0.8746(6) & -0.9072(5)\\
& atomic & -0.8581(3) & -0.9673(4) & -0.9980(4) & -1.019(4) & -1.0286(3) & -1.0419(6) & -1.0556(4) &-1.0724(2) \\
& rel. &   -0.8573(3) & -0.9665(3) & -1.0033(3) & -1.0185(6) & -1.0333(2) & -1.0707(3) & -1.0547(4) & -1.0720(3) \\
& bi-atomic & -0.8558(3) & \cellcolor{green}-1.0437(2) & \cellcolor{green}-1.1051(2) & \cellcolor{green} -1.1255(2) & \cellcolor{green} -1.1327(1) & \cellcolor{green} -1.1370(2) & \cellcolor{green} -1.1395(1) & \cellcolor{green}-1.1417(1) \\
& rel. & -0.8553(4) & \cellcolor{orange}-1.0433(1) & \cellcolor{orange} -1.1048(2) & \cellcolor{orange} -1.1255(1) & \cellcolor{orange} -1.1322(2) & \cellcolor{orange} -1.1369(1) & \cellcolor{orange} -1.1392(1) & \cellcolor{orange}-1.1417(1) \\
\hline
\end{tabular}
\caption{Energies per atom in units of Ry for the different crystal structures: square, triangle, atomic graphene (graphene-a) and molecular graphene (graphene-b) and underlying orbitals in the Slater determinant of the VMC wave function: plane wave (pw), DFT, atomic, and bi-atomic orbitals. In each line below the orbitals types pw, atomic, and bi-atomic, we give the energies of the structural relaxation (rel.) based on the corresponding orbital. The green (orange) fields mark the lowest energies before (after) structure relaxation.}
\label{energytable}
\end{table*}
\endgroup

Based on the generalized forces given above, we have continued to optimize the protonic structure for plane-wave, atomic, and bi-atomic orbitals, starting from the aforementioned crystal structures investigated here. 
The relaxation using the DFT orbitals, whereas possible in principle, has not be taken into consideration in this work, as it complicates enormously the wave function minimization process.
Our results after relaxing the positions of the protons are shown in Fig.~\ref{fig:StructureRelaxation_energies}. 
The protonic structure optimization results in a lower energy for most of the configurations and trial wave functions. 
Energies after relaxation become less sensitive to the initial structure and group together depending mainly on the choice of the underlying Slater orbitals.

The molecular graphene-like structure remains the favored low density/pressure phase, and it is best described by the bi-atomic orbitals. 
Energies in this molecular phase are unaffected by the relaxation within our statistical uncertainties. 
This is in contrast to our results at high densities, $r_s \lesssim 1.1$, where relaxation lowers significantly the energy of the triangular structure with plane-wave orbitals, the favored ones between the orbitals used for the structure relaxation.
This might indicate that the ground state of the atomic phase is not likely to be described by a simple triangular crystal structure, but contains more atoms in the unit cell, similar to the high pressure structures predicted in 3D \cite{PhysRevLett.106.165302,PSSB:PSSB200880546}. 
Another possibility is that the number of atoms used in our simulation is compatible with the real ground state structure, but the relaxation process falls in a local minimum.

Besides the structure of hydrogen, its conductivity is certainly the most interesting property. 
Here, instead of attempting a direct calculation of the conductivity \cite{PhysRevLett.103.256401}, we simply deduce metallic or insulating behavior from the localization properties of the ground state wave function. 
Within VMC, the localization of the reduced single particle density matrix directly reflects the character of the orbitals inside the Slater determinant \cite{PhysRevLett.114.105701,Pierleoni03052016}. 
Since the molecular graphene structure is described by localized bi-atomic orbitals, and the triangular structure with extended DFT/pw orbitals, the metallization transition occurs together with the structural phase transition around $r_s \approx 1.1$ within our description.


Finally, let us provide a rough estimation for the pressure necessary to reach metallization in 2D. 
The two-dimensional pressure as obtained from an approximated Maxwell construction is estimated to be around $54~\text{N}/\text{m}\simeq 1.0 \cdot 10^3 a_0 \,\text{GPa}$ with an error of a few percent due to the uncertainty of the Maxwell construction. The orbitals currently used in the relaxation are likely to favour the molecular phase, so that the pressure should present an upper bound for metallization in strictly 2D hydrogen.



In conclusion, we have introduced a novel general algorithm that can be used for simulations in the spirit of Car-Parrinello, with the major difference of replacing density functional theory with the more accurate quantum Monte Carlo methods for describing the electronic structure.
Whereas this original approach might open up a new generation of ab-initio simulations of higher accuracy, we have confined ourselves to the case of $T~=~0~\text{K}$, where the dynamics of the protons simply leads to a structure relaxation. 
We have shown that it is possible to use this method for geometrical optimizations, and we have applied it to investigate high pressure $2$-D hydrogen structures where protons are confined within a plane and electrons are free to move in $3$-D. 
Our simulations indicate metallization at $r_s~\approx~1.1$ and at a pressure of $1.0 \cdot 10^3~a_0~\text{GPa}$, together with a structural transition from a molecular lattice to an atomic phase. 
Whereas the geometrical optimization confirms the molecular graphene structure for the insulating phase, the structure of the metallic phase is ambiguous. 
However, there are good circumstantial evidences that all of the considered starting configurations can be excluded with reasonable certainty for the atomic phase. 
This either means that we have not considered a number of atoms compatible with the ground state unit cell of the atomic phase, or that a minimization technique better suited for finding a global minimum should be adopted for addressing this specific question. 
Finally, we would like to mention that using DFT orbitals  in the structural relaxation is likely to further lower the energy in the metallic state so that metallization might occur at a slightly lower pressure.

\section*{Additional Material} 
\label{sec:appendix}

\subsection{Ewald summation in quasi-2D layers} 
\label{sub:appendices}
In our study, we have considered a $2D$ layer with periodic conditions on the $x$ and $y$ axis.
However, the electrons were allowed to move in a $3D$ space.
Such a peculiarity requires some corrections in the Ewald summation.

In particular, when summing over all $\mathbf{k}$-vectors in the long-range term, one should consider the limit $L_z \rightarrow \infty$ and therefore $\Delta k_z \rightarrow dk_z$. The sum over all $k_z$ needs to be substituted by an integral:
\begin{equation}
   \sum_{k_z} \frac{e^{-\frac{k_z^2}{4 \alpha}}}{k_x^2 + k_y^2 + k_z^2} e^{i k_z {r_{ij}}_z} \rightarrow \int_{-\infty}^{\infty} dk_z \, \frac{e^{-\frac{k_z^2}{4 \alpha}}}{k_x^2 + k_y^2 + k_z^2} e^{i k_z {r_{ij}}_z}\, . \label{eq:int_kz_ewald}
\end{equation}

This integral can actually be computed analytically.
By using the substitution
\begin{equation}
   \frac{1}{k_x^2 + k_y^2 + k_z^2} = - \int_{0}^{\infty} dt \, e^{- \left( k_x^2 + k_y^2 + k_z^2 \right) t} \, ,
\end{equation}
one can compute the integral in Eq. \eqref{eq:int_kz_ewald}, obtaining
\begin{widetext}
\begin{equation}
   \frac{\pi e^{\frac{1}{4 \alpha} k_x^2 + k_y^2 - \sqrt{k_x^2 + k_y^2} |{r_{ij}}_z|}}{2 \sqrt{k_x^2 + k_y^2}} \left[ 2 + e^{2 \frac{1}{4 \alpha} \sqrt{k_x^2 + k_y^2} + |{r_{ij}}_z|} \erfc{ \frac{2 \frac{1}{4 \alpha} \sqrt{k_x^2 + k_y^2} + |{r_{ij}}_z|}{2 \sqrt{\frac{1}{4 \alpha}}} } - \erfc{ \frac{-2 \frac{1}{4 \alpha} \sqrt{k_x^2 + k_y^2} + |{r_{ij}}_z|}{2 \sqrt{\frac{1}{4 \alpha}}} } \right] \label{eq:z_ewald}
\end{equation}
\end{widetext}

The long-range term in the Ewald summation then reads
\begin{equation}
   \sum_{i,j} \sum_{k_x, k_y} - \frac{4 \pi}{\Vol} e^{-\frac{k_x^2 + k_y^2}{4 \alpha}} e^{i \left( k_x {r_{ij}}_x + k_y {r_{ij}}_y \right)} \mathcal{F}(k_x,k_y,\alpha,{r_{ij}}_z)
\end{equation}
where $\Vol$ is the volume of the system, $\mathcal{F}$ is the expression in Eq. \eqref{eq:z_ewald}, and the sum over $k_x, k_y$ does not include the case $k_x=k_y=0$.


\begin{figure}[htbp]
  \centering
    \includegraphics[width=0.45\textwidth]{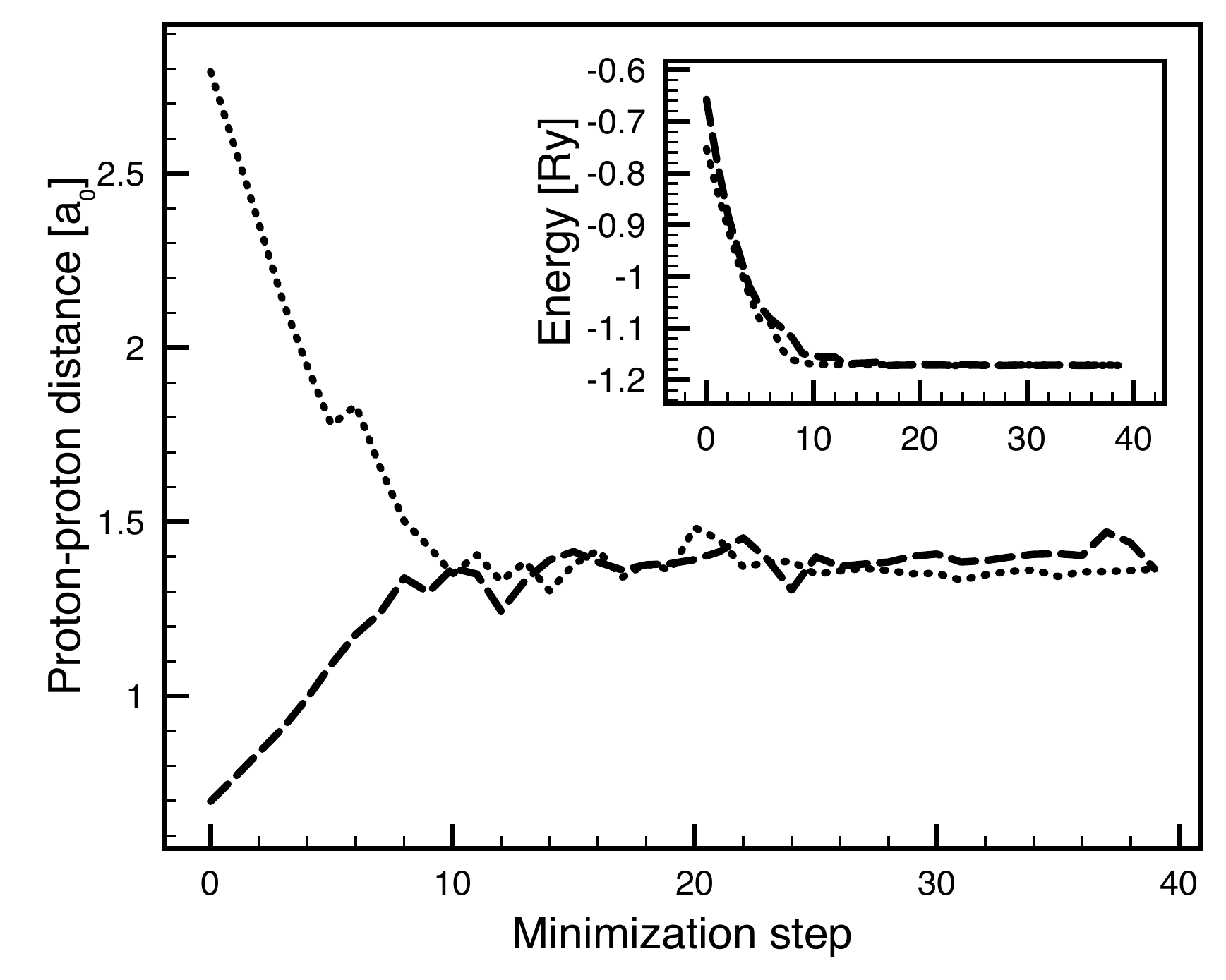}
  \caption{Structure optimization algorithm applied to the simple \Htwo molecule. We used the bi-atomic wave function, and two different starting distances for the protons. The inset shows the corresponding evolution of the variational energy.}
  \label{fig:H2_binding}
\end{figure}

\subsection{Code and Algorithm Reliability Check} 
\label{ssub:code_and_algorithm_reliability_check}
Whereas the VMC part was used before \cite{paper:calcavecchia_kuehne} and it is therefore known to provide reliable results, the structure optimization algorithm was new.

In order to check that both our novel algorithm and code work as expected, we have applied it to a very simple case: The \Htwo binding.
The results of our test simulations are presented in Fig.~\ref{fig:H2_binding} and demonstrated the reliability of our calculations.


\subsection{Finite size effects} 
\label{sub:finite_size_effects}
Finite-size effects are known to play a crucial role in solid state physics, and in particular in conductive materials. In our study, we have accounted for them by means of the TABC when using the plane-wave wave function, but not when using the atomic and bi-atomic ones. This approximation is justified as long as the simulation box is big enough for containing such localized wave functions. However, when dealing with very high densities, it is legitimate to wonder if such an approximation is valid or not.

We verified the validity of our approximation by explicitly computing the kinetic energy for the atomic orbital for a finite and infinite simulation box.
In particular, we have considered the atomic graphene-like structure at $r_s=1$, where $L_x=27.99~\text{Bohr}$, $L_y=16.16~\text{Bohr}$, and $\Catm=0.421$. We found out that our energies are reliable up to $\sim~10^{-3}~\text{Ry}~\simeq~10^{-2}~\text{eV}$. This uncertainty is much larger than the statistical error inherited from the Monte Carlo integration.



\begin{acknowledgments}
F.C. and M.H. thank the NanoScience Foundation for support and acknowledge discussions with David Ceperley, T. D. K\"uhne for allowing us to access the Mogon HPC which has been used for most of the numerical calculations.
\end{acknowledgments}

\bibliographystyle{apsrev}
\bibliography{references}

\end{document}